*Original Article*

# Autonomous Threat Detection and Response in Cloud Security: A Comprehensive Survey of AI-Driven Strategies

Gaurav Sarraf[1], Vibhor Pal[2]
[1,2]Independent Researcher

**Abstract -** Cloud computing has changed online communities in three dimensions, which are scalability, adaptability and reduced overhead. But there are serious security concerns which are brought about by its distributed and multi-tenant characteristics. The old methods of detecting and reacting to threats which are mostly reliant on fixed signatures, predefined rules and human operators are becoming less and less effective even in the advanced stages of cyberattacks of cloud infrastructures. The recent trend in the field of addressing these limitations is the creation of technologies of artificial intelligence (AI). The strategies allow independent protection, anomaly detection, and real-time analysis with references to using deep learning, machine learning, and reinforcement learning. Through imbuing AI with a constantly-learning feature, it enables the intrusion detection system to be more accurate and generate a lesser number of false positives and it also enables the possibility of adaptive and predictive security. The fusion of large-scale language models with efficient orchestration platforms contributes to reacting to the arising threats with a quicker and more precise response. This allows automatic control over incidences, self-healing network, and defense mechanisms on a policy basis. Considering the current detection and response methods, this discussion assesses their strengths and weaknesses and outlines key issues such as data privacy, adversarial machine learning and integration complexity in the context of AI-based cloud security. These results suggest the future application of AI to support autonomous, scalable and active cloud security operations.

## 1. Introduction

Cloud computing is an essential aspect of the modern digital infrastructure that allows on-demand access to common resources, including storage, applications, and computing power [1][2]. It has been relevant to individuals, governments, and businesses due to its abilities, elasticity, and low-cost [3]. Its flexibility, scalability, and cost effectiveness have made it significant to both individuals, governments, and even businesses. Enterprises are transferring critical data and services to the cloud without considering their security vulnerabilities [4][5]. The vulnerability of cloud infrastructure to attacks is because the dynamic nature of the cloud environment, multi-tenancy, and distributed architecture provide hackers with greater opportunities to compromise it [6][7]. Cloud computing resilience has therefore played a vital role in the need to create trust and credibility in virtual ecosystem.

One of the pillars of cloud security becomes threat detection [8]. Threats that may negatively affect the privacy, availability, and integrity of information are data insider threat, malware implantation, denial-of-service attacks, and unauthorized access [9]. The ever growing complexity of cloud-based attacks is straining the traditional approaches of detection which employs fixed rules and signature matching [10]. Moreover, the scale and variety of cloud traffic make it more difficult to monitor it manually because it requires resources and is likely to cause errors [11][12]. This has been in development of the futuristic structures of detection capable of identifying anomalies, reducing false positives and delivery of live situational awareness.

Threat response is also significant in order to reduce the damage of cyber-incidents [13][14]. Cloud security threat detection and response do not merely mean the capacity to detect malicious activity but beyond this capacity it means the capacity to contain and minimize the impacts of the malicious activity within a short time [15][16]. The traditional response models are likely to involve human intervention that is not always fast in the mass attack, or real-time attack [17][18]. Therefore, automated response mechanisms are on the rise, enabling faster responses by isolating affected resources, introducing security rules, and preventing lateral infection from attacks [19]. One of the resolutions required is automation with the aim of establishing resilience and agile cloud defines infrastructures.

AI has emerged as a pillar in boosting the ability of detecting and responding to threats in clouds [20][21]. Under AI-based systems, with the help of ML, DL, and RL, vast quantities of heterogeneous data on clouds are processed and novel attack vectors are discovered, and the independent triggering of a defensive reaction is created [22][23]. Outside of reactive security, AI enables predictive analytics, adaptive policy enforcement, and continuous self-learning, making the system resilient to changes in threats [24]. Combining information related to logs, user actions and network traffic, AI helps to minimize false positives and increase situational awareness [25]. Furthermore, response systems based on AI focus more and organize countermeasures in real-time, which opens the way to scalable, self-surging and autonomous cloud security systems.

### 1.1. Structure of the Paper
The paper structure is as follows Section II The Threat Landscape and Security in the Cloud. Section III AI enable threat detection in cloud Section IV. Strategies for autonomous threat response are discussed in Section V of the literature review, and in Section VI, conclusions and future research are discussed.

## 2. Cloud Security and Threat Landscape
Computing in the cloud has become essential for contemporary digital services due to its scalability, adaptability, and low operating costs. A complex danger landscape is introduced by its dispersed, multi-tenant nature, and specific security procedures are required. The rise of cloud computing has revolutionized how companies manage their information technology by making shared computer resources that were previously unavailable available online at all times (Figure 1).

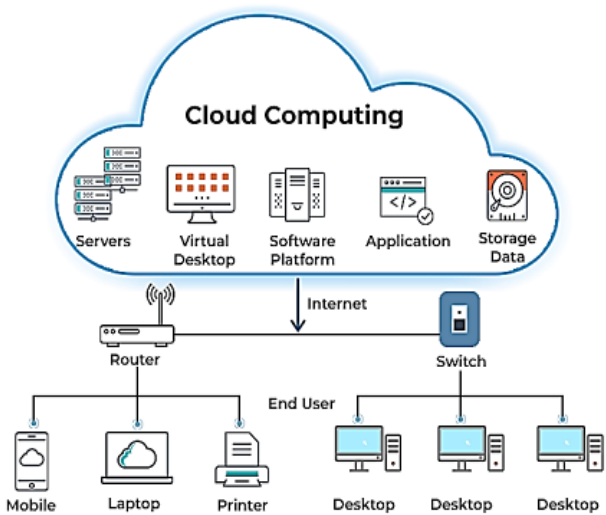

**Fig 1: Cloud Computing Architecture**

This paradigm of computing makes use of distant data storage, processing, and access via an internet connection. [26] It incorporates storage, apps, software platforms, virtual desktops, and servers.

### 2.1. Cloud Computing Models and Architectures
Cloud computing offers several service models and deployment architectures that allow users to access shared computer resources on demand. Delivery, management, and consumption of services inside the cloud ecosystem are defined by these models.

#### 2.1.1. Service-Based Model
There is a wide variety of services offered by cloud computing at the moment, but the three main ones are SAAS, IaaS, and PaaS (Figure 2), which is also called the Cloud service model.

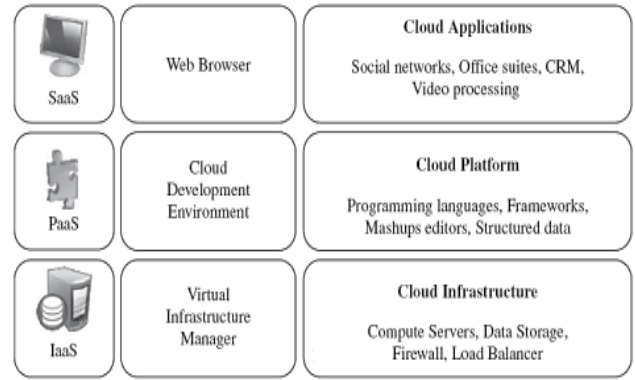

**Fig 2: Cloud Service Model**

#### 2.1.2. Infrastructure as a Service (IaaS)
Computing resources are primarily comprised of hardware and software components [27]. The bedrock of any computer system is this. In the cloud, these services are provided to end users via Infrastructure-as-a-Service[28]. So, IaaS basically removes the need for users to pay for these services.

2.1.2.1. Example of (IaaS): Microsoft Azure, AWS, Cisco Metacloud.
- *Platform as a Service (PaaS): SaaS, leveraging APIs, programming languages, and development middleware, enables subscribers to develop their own applications without the need to establish or configure the development environment [29]. Companies that offer Platform as a Service include Google App Engine, Microsoft Azure, and Salesforce.com.*
  **Example of (PaaS):** AWS Elastic Beanstalk, Apache Stratos, Google App Engine, Microsoft Azure.
- *Software as a Service (SaaS): SaaS models allow users to subscribe to or pay for cloud-based software and services on an as-needed basis. The only software required to access cloud-based SaaS applications is a web browser or thin client [30]. This reduces load on end-user hardware and also provides the capability of centralized program control, deployment and maintenance.*
  **Examples of (SaaS):** Microsoft Office 365, Salesforce, Cisco WebEx, and Google Apps.
- *Deployment Model:* The ambiguity of the three likely solutions is a challenge to online businesses in deciding whether to invest heavily in private, public or hybrid cloud computing [31].

**Public Cloud**
The public cloud model is one of the frequent applications of the cloud. The applications of this type of cloud are common in web applications, sharing files, and storing data that are not confidential [32]. Public clouds are the solution when developing software.

**Private Cloud**
There are several names to which the method of deploying a private cloud is identified: internal, corporate, and private. The resources of an organization in a private cloud are unique. Businesses will usually prefer to have their hardware stored in their in-house data centre.

**Hybrid Cloud**
Hybrid clouds combine both the public and private clouds. Their architecture focuses on the smooth movement of data and applications, and also smooth communication between the two platforms [33]. The two types may be necessary in organizations of any size and in any industry and this is the best alternative that suits groups which fit into the category..

*2.1.3. Architectural Foundations of Cloud Systems*
The following are the key players of the architectural foundation of cloud systems:

2.1.3.1. Virtualization and Containerization
Cloud computing is based on the technologies of virtualization and containerization that can be used and provide efficient utilization of resources, scalability, and isolation.

**Virtualization**
- Cloud computing is based on two basic technologies that enable efficient use of resources, scalability, and isolation, which are virtualization and containerization.
- A hypervisor manages the use of virtual machines by different workloads and this allows multiple workloads to effectively use the same physical infrastructure. Popular hypervisor vendors are VMware, Hyper-V and KVM.
- It is more flexible, isolates faults and hardware independent but generally has a greater overhead than containers.

**Containerization**
- Software-as-a-service (SaaS)
- The containers take less time to start and require fewer resources compared to virtual machines (VMs) as they share the host operating system kernel.

2.1.3.2. Multi-Tenancy
Multi-tenancy in cloud environments has significantly enhanced the scalability, efficiency, and cost-effectiveness of SaaS applications [34]. By sharing resources across multiple tenants, SaaS providers can reduce per-customer operational costs while maximizing infrastructure utilization. Multi-tenancy also streamlines software deployment and maintenance.

2.1.3.3. Elasticity and Scalability
Elasticity is the capacity of cloud systems to optimize costs and avoid over-provisioning by dynamically allocating or releasing resources in response to real-time demand. However, scalability refers to a system's ability to handle increased workloads, whether by increasing the capacity of individual machines or the overall system.

*2.2. Threats and Vulnerabilities in Cloud Environments*
A threat is something that could trigger an event that could damage a system or an organization. An asset or system is vulnerable if it has any openings that a threat could use to compromise it. A threat actor can launch an attack by exploiting a security hole [35]. A comprehensive literature review identified numerous risks and vulnerabilities in cloud computing. Following is an in-depth discussion of them.

*2.2.1. Data Breaches*
An incident of data breach happens when someone or some group without authorization obtains, copies, or transmits private, sensitive, or otherwise privileged information about a person or company. The most serious and pervasive danger in cloud computing is the possibility of a data breach. Threat actors' deliberate actions or even inadvertent mistakes can lead to data breaches.

*2.2.2. Data Loss*
Data corruption or unavailability can occur for many reasons. Some of these reasons include natural disasters like floods or earthquakes, and on the other hand, human error like cloud administrators erasing data by accident, hard drive failure, power outages, or malware infestations.

*2.2.3. Malicious Insiders*
The greatest danger comes from hostile insiders, who provide perhaps the most catastrophic threat. Potential insider threats can come from anyone with ties to the company, including former employees, system administrators, third-party contractors, and business partners.

*2.2.4. Denial of Service*
The availability of a system is affected by a denial-of-service (DoS) attack, as seen in Figure 3, demonstrating that a single-machine denial-of-service attack can be reduced.

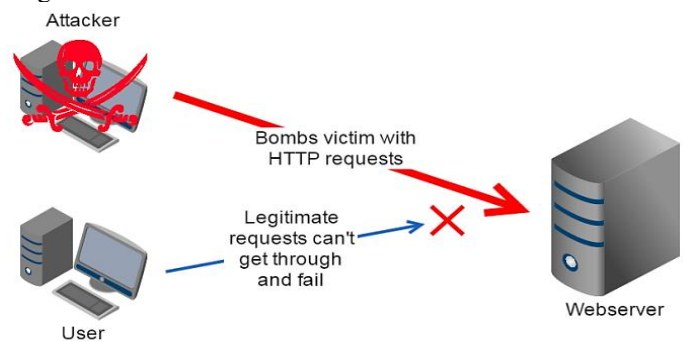

**Fig 3: Denial of Service**

DoS and DDoS assaults are simple to launch, particularly with the help of a botnet.

*2.2.5. Vulnerable Systems and APIs*
The public can access cloud application using its Application Programming Interfaces (APIs). After successfully exploiting a cloud API, an attacker can gain extensive access to cloud resources [36]. Vulnerabilities in the underlying operating system, which are flaws in the system programs that an attacker may use to take over the system or prevent the service from running, could exist at the PaaS level.

*2.2.6. Account Hijacking*

Services that are quite noisy bring a fresh danger to the scene of service or account hijacking. When an unauthorized third-party gains access to a valid user's account and uses it for malicious purposes, this is known as account hijacking.

*2.2.7. Advanced Persistent Threats*

The objective of an Advanced Persistent Threat (APT) assault is to get access to a company's cloud infrastructure and steal data once the attackers have gained a foothold. Threat actors with advanced persistence can gain access to cloud services through several means, including spear phishing, direct hacking, attack code on USB drives, and network penetration.

**2.3. Conventional Threat Detection and Response in Cloud Security**

Cybersecurity has traditionally relied on signature-based, rule-driven, and manual intervention approaches for threat detection and response in the cloud [37]. Antivirus software, firewalls, and other intrusion prevention and detection systems, as well as manual incident response procedures, fall under this category. Here are a few things that can be detected:

*2.3.1. Signature Based Detection*

Conventional security measures identify harmful actions by comparing observed patterns to previously recorded attack signatures or malware definitions. While effective against familiar threats, they struggle with zero-day attacks and polymorphic malware

*2.3.2. Rule-Based Defense Mechanisms*

Predefined rules are used to restrict unauthorized access by access control lists, firewall policies, and static configurations. They are however less effective in dynamic clouds where the workloads keep changing and thus they are very rigid.

*2.3.3. Manual Incident Response*

Security teams have to search, triage, and correct incidents manually. This often consumes resources, time and prone to human error especially when it comes to large scale or real time attacks.

## 3. AI-Enabled Threat Detection in Cloud Security

**3.1. Machine Learning Approach**
Machine learning methods are based on the construction of an explicit or implicit model that make it easier to classify the patterns being studied [38]. Tagged data is also necessary to train the behavioral model and is unique to these schemes and consumes resources heavily:

**3.2. Supervised Learning**
Train supervised learning algorithms on labelled datasets, where each data point is linked to a particular output. The cybersecurity industry heavily relies on these algorithms for categorization tasks, like differentiating between safe and dangerous traffic. Commonly used supervised learning algorithms.

- **Decision tree:** Decision trees are well-liked because they are easy to understand and use. In order for them to classify input data, they recursively divide it into subsets according to feature values. Ultimately, this process culminates in a judgment node.
- **Naïve Bayes:** The probability value of the Naïve Bayes algorithm is determined using a simple probabilistic technique that is employed in classification.

*3.2.1. Unsupervised Learning*

Unsupervised learning algorithms are perfect for discovering new or undiscovered dangers since they don't need labelled data. Using structures and patterns in the data, these algorithms determine if a user is acting normally or abnormally. Its main applications are in feature reduction and clustering two main methods for dimensionality reduction and clustering.

- **Clustering:** One way clustering methods function is by dividing up the collected data into groups based on some distance or similarity metric. One typical method is picking a single point from each cluster to serve as a representative.
- **PCA:** Data dimensions are diminished by Principal Component Analysis (PCA). To get a better understanding of PCA, two axes are used to plot the data when it is displayed on a graph.
- **Autoencoders,:** a type of neural network used for unsupervised learning, have also gained popularity in anomaly detection

*3.2.2. Reinforcement Learning*

A potent AI method for handling dynamic and complicated situations, like SaaS systems, is reinforcement learning, or RL. Because of changes in infrastructure, application load, and user behavior, resource consumption varies in these settings. While traditional rule-based or static techniques are unable to effectively adjust to these changes.

- **Q learning:** A model-free approach to learning the value function in situations when the system dynamics are uncertain.
- **Deep Q-Networks (DQN):** Integrate Q-learning and deep neural networks to manage expansive state action spaces; a valuable asset in intricate SaaS systems.

**3.3. Deep Learning for Complex Pattern Recognition**
Deep learning is a game-changer for making sense of multidimensional datasets and large-scale time series, revealing complex patterns and long-term connections.

- **Bayesian networks:** Bayesian networks are models that encode the probability relationships among relevant variables. Common applications of this method include intrusion detection in conjunction with statistical techniques.
- **Long short-term memory (LSTM):** LSTMs are a kind of RNNs tailored for use with sequence data, including text, audio, and time series.
- **Neural network:** Anomaly intrusion detection, flexibility, and adaptability to changes in the environment are some of the areas that have embraced neural networks [39]. Several applications have made use of this detection method, including user profile creation, command sequence prediction, and the identification of intrusive traffic pattern behavior.

*3.4. Anomaly and Intrusion Detection System*

Cloud security isn't complete without intrusion and anomaly detection systems (IDS), which keep an eye on system operations and network traffic for signs of hostile activity[40]. The first database keeps track of the current network profiles across time, and the second keeps track of the statistical profiles that have been taught in the past. Information Detection Systems heavily on the knowledge-based approach [41]. In this case, the input data labels are established using a set of rules, which consists of two steps. Probe, DOS, U2R, and R2U labels are used by some anomaly detectors to categorize aberrant data, which helps to describe the type of assault that was identified.

Cyberattack using signatures known attack signatures and framework vulnerabilities are the basis for this detection system's ability to spot suspicious activity [42]. One method for identifying potential abuse is by comparing observed behavior with signatures of known attacks.

*3.5. Challenges of AI-Powered Threat Detection in Cloud Environment*
- **Complexity of Integration:** Existing cloud architectures can be resource-intensive and hard to implement AI systems. It calls for a high level of knowledge in artificial intelligence and cloud computing, neither of which are necessarily present in any organization.
- **Data Privacy and Security:** AI is employed in cybersecurity to process substantial quantities of potentially sensitive data. It is not an easy task to guard the confidentiality of this information, especially when facing tough regulatory standards like GDPR or HIPAA.
- **Dependency on Data Quality:** The quantity, diversity, and quality of data used to feed AI models have a significant influence on the performance of the models. Poorly performing models that are either unable to predict dangers or, even worse, error-prone, can be the result of bad or biased data.
- **Adversarial Attacks:** AI systems can learn to recognize a threat, and so can the assailants learn how to bypass AI systems. Adversarial machine learning is a relatively recent field of research that relies on manipulation of data to deceive AI algorithms, and it may lead to security breaches.
- **High Initial Costs:** The original configuration, involving creating and incorporating AI models, may prove expensive, yet AI may ultimately prove cost-effective.

## 4. Autonomous Threat Response Strategies

Autonomous threat response strategies represent a paradigm shift in cloud security by enabling systems to detect, analyze, and neutralize threats with minimal human intervention [43]. These strategies combine automation, artificial intelligence, and orchestration platforms to ensure faster and more accurate responses to evolving cyberattacks some key point discussed are given below:

*4.1. Automated Incident Handling and Mitigation*

Cyberattacks nowadays are so large and complicated that conventional incident response systems aren't able to handle them. AI enhances incident response by automating the activities related to identification, assessment, and mitigation of security incidents.

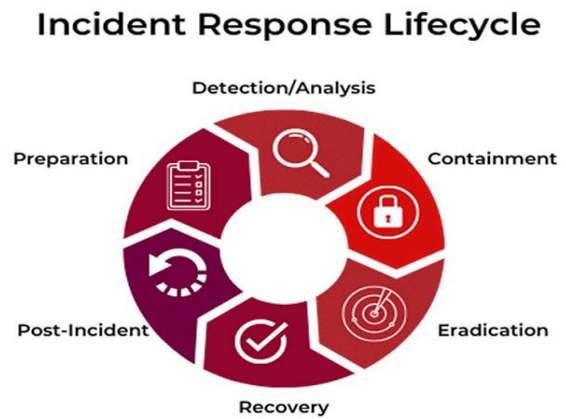

**Fig 4: Incident Response Lifecycle in Threat**

This strategy as depicted in Figure 4 ensures that the methods of responding to incidents are strong. There are, however, security risks associated with the usage of AI. Some of the elements employed in the process of regulation are safety measures and contingency plans.
- **Containment, Eradication & Recovery:** Containment operations refer to such operations that will reduce security incidents. They can be band-aid measures which only short-term reduce the hazards, or they can be comprehensive in eliminating them.
- **Threat Detection and Classification:** The capability to recognize and categorize novel threats based on massive amounts of unstructured data, including chat logs, security logs, and emails, is known as incident response. LLMs in transformer models, e.g. BERT and GPT, can be used to identify patterns and anomalies, which could be indicative of security events. These LLMs excel at NLP tasks.
- **Real-time Response Automation:** A range of incident response controls, which accelerate containment and recovery activities and minimize the burden on human intervention [44]. A threat can cause planned procedures, like closing down rogue

IP addresses and isolating infected systems, should a threat be detected. Such capabilities are also supported by the ability of the Security Orchestration, Automation, and Response (SOAR) to introduce the capabilities of LLM.
- **Improving Post-Incident Analysis:** post-occurrence investigation. Based on the data on the historical events, LLMs can disclose the motive of a breach and ways used by the attackers.

### 4.2. Adaptive and Self-Healing Mechanisms
Dynamic Network Segmentation: A dynamic network segmentation of a self-healing network is one of the most notable transformations of the traditional notion of a static network design, which is represented by highly reconfigurable segments synchronized by analyzing network behavior and detecting danger in real-time [45].

**Automated Isolation and recovery:** The automated isolation and recovery procedures form a highly significant component of self-healing networks and are what enable the latter to withstand ransomware intrusions. In the event of detection of something abnormal by the monitoring systems in the network.

### 4.3. Policy-Driven and Real-Time Defense Model
Cloud security policies and real-time models of defense emphasize the application of dynamic security policies and offer real-time response to changing threats. These models are effective because they incorporate artificial intelligence and pre-defined organizational policies, to make sure that cloud environments are compliant, resilient, and adaptive to attack conditions.

#### 4.3.1. Policy-Based Security Enforcement
A policy-based security model that can be adjusted to meet the specific security requirements of each use case, allowing for a security architecture that is both flexible and scalable over the device's lifetime [46]. To enforce policies, either an external hardware item or a software component like SELinux might be used. The device can be protected from any attempts at re-purposing or exploit by using a hardware-based intrusion mitigation approach. Listed below are the components of the policy enforcement engine:
- **Hardware:** Enforcing the specified security standards for every slave while keeping an eye on system-level communication between memory peripherals and data.
- **Software:** Recognizing unusual activity and verifying application access limits.

#### 4.3.2. Real-Time Threat Detection and Response
The ability to detect and respond to cyber threats in real time, reducing disruption and harm, is known as real-time threat detection [47]. The ability to make purchases is facilitated by the constant surveillance of the systems, networks and user activities. There are several technologies required in order to facilitate the detection of threats in real time:
- The SIEM (Security Information and Event Management) system is capable of detecting IOCs and threats across the enterprise in real-time by real-time analytics and correlation processes.
- The Endpoint Detection and Response (EDR) technology detects attacks bypassing conventional control barriers using endpoint activity monitoring and behavioral analysis [48].
- Security Orchestration, Automation and Response (SOAR) is a framework that can help in the creation of playbooks, which when activated can automatically initiate processes to contain risks.

#### 4.3.3. Adaptive Policy Updating
Adaptive Policy Updating applies AI and threat intelligence to then proceed to continually optimize security policies based on the evolving attack patterns. Rather than depending on rules that are static, cloud systems dynamically modify access controls, firewall settings and anomaly thresholds in real time, to combat created threats, like zero-day exploits [49]. The effectiveness of the policy enforcement in the constantly evolving cloud environments is maintained by the continuous learning process that reduces the number of false positives and ensures resilience.

## 5. Literature Review
This section presents the existing literature on autonomous threat detection and response mechanisms that are enabled by AI in cloud security. Table I presents a literature review of the literature that presents different cloud-based AI-based threat identification and incident response models and issues of implementing the model:

Pitkar (2025) explores the present condition of cloud security automation, paying attention to how it can create a state of symmetry between the ability to detect and actually react to threats. This paper presents the major technologies: Security information and event management (SIEM), Extended Detection and Response (XDR) and Security Orchestration, Automation, and Response (SOAR) platforms by examining changes that have occurred in the market in recent years and examining the latest technological trends. Artificial intelligence and machine learning integration have changed the nature of these systems as they are now able to detect threats in real-time and provide automatic response systems. The study discusses practical examples and points out that organizations that use automated security systems have shown a shortened incident response time and security breaches. Nonetheless, there are still issues of complexity of integration and equivalence between automation and human experience [50].

Akinloye et al. (2024) discovered that AI-based systems enhance threat detection and response speed and accuracy by a significant margin. AI has the ability to automate a lot of processes, which might make human analysts obsolete and allow for quicker threat mitigation. The fact that AI is applicable in so many different fields is further proof of its versatility and resilience in the face of changing dangers. Finally, progress in cybersecurity for networks supporting national infrastructure has been made with the advent of AI-driven threat detection and response systems. Thus, these technologies help strengthen national infrastructures as a whole by making it easier to detect and counteract cyberattacks [51].

Mahajan et al. (2023) Cloud security problem identification and problem solving! Digital forensics can be enhanced with AI models to make the incident triage and problem detection faster. Improved cloud security, early threat detection, efficient use of resources, and compliance with all applicable rules and regulations  To better handle cloud security issues in a constantly changing digital world, it is possible to employ advanced AI models, automated incident response, human-machine collaboration, adversarial machine learning, compliance and legal considerations, and cross-cloud security.  A more proactive and determined strategy [52].

Sidhu (2023) Network Intrusion Detection (NID) is the cloud computing security measure that assists in detecting intrusions on cloud services and cloud infrastructure.  One of the elements of the operation of NID is the monitoring of network traffic and its analysis in case of malicious activity. Data leakage, DDoS attacks, malicious traffic, unauthorized access, and illegal access are just some of the hazards that it can identify.  With NID's early risk detection capabilities, cloud systems may be better protected from damage and data breaches.  Another perk is that NID systems are cloud-ready and secure.  The ability to identify harmful actions makes NID an ideal solution for use in cloud environments [53].

Ademilua and Areghan (2022) discussed cloud security frameworks powered by AI. Detecting and mitigating complex, ever-changing threats is becoming more difficult for traditional security systems in cloud infrastructures. Tools that improve threat detection, policy automation, and data protection automation analyze a variety of artificial intelligence techniques, including federated learning, DL, NLP, supervised and unsupervised learning, the impact of interpretability tools, and the importance of regulatory compliance [54].

Sailesh Oduri (2021) developed machine learning models that may anticipate and detect irregularities that may indicate possible security breaches by combining threat patterns with real-time monitoring.  Significant findings indicate that the threat of such systems is more efficiently identified by AI-driven systems and fewer false positives are generated, which contributes to their increased reliability. The models showed a considerable increase in speed and accuracy of threat identification on a variety of cloud platforms as compared to more traditional methods. This article delves into the potential of AI to transform cloud security, outlines the challenges encountered during deployment (such as data privacy concerns), and suggests research paths to enhance AI capabilities in this domain [55].

Patel et al. (2020) Data storage and online business applications are examples of the kinds of computing services made possible by this new technology. Organizations can save money, have access to data and information security, and more by moving their operations to the cloud, which allows for a distributed work environment. Cybercriminals are constantly developing new methods to breach cloud storage and steal critical information as more companies transfer their operations to the cloud. Because of the shift from on-premises to cloud computing, many security concerns have surfaced for both customers and service providers. As a result of the wide variety of technologies used to deliver various services by reliable cloud providers over the Internet, new security risks have emerged [56].

Torkura et al. (2019) One method to identify and mitigate security incidents is to implement recovery and dynamic snapshotting procedures. The second method improves upon the first by automatically correlating the generated alerts with the cloud event log to retrieve useful information for incident response. Malicious actions are thus examined, uncovered, and eradicated. Together, the suggested methods for protecting AWS and GCP against the aforementioned security threats operate in near real time. Cloud attacks, both static and dynamic, were used to test methodologies. While AWS and GCP both have Mean Times to Respond of 25 and 90 minutes, respectively, the average Mean Time to Detect for both providers is 30 seconds [57].

Al-Mohannadi et al. (2018) A growing number of cyber-attacks have had devastating effects on both individuals and businesses. IDS and IPS are two of the several technologies that have emerged to deal with cyberattacks. There an overwhelming amount of false positive alarms generated by these solutions. In order to successfully filter the massive volume of alerts, identify actual affirmative attacks, and respond according to the incident response rule, the IDS tool results should be controlled by an intelligent human. The IT staff must be well-versed in the ins and outs of IDS, IPS, and managing incidents [58].

**Table 1: Comparative Analysis of AI-Driven Cloud Security in Threat Detection**

| Author(s), Year | Focus Area | Key Findings | Challenges | Future Work |
|---|---|---|---|---|
| Pitkar (2025) | Cloud security automation, especially the integration of SIEM, XDR, and SOAR systems enhanced with | AI and ML significantly enhance real-time threat detection and automated incident response. Automated security solutions improve incident response times and reduce | - Complexity involved in integrating multiple automation platforms. - Maintaining balance between automated systems and human decision-making. - Need | - Develop methodologies to streamline platform integration. - Research on human–AI collaboration models. - Improve interpretability of AI-driven decisions in security operations. |

| | AI/ML technologies | the likelihood of breaches. | for skilled personnel to manage AI-driven tools. | |
|---|---|---|---|---|
| Akinloye et al. (2024) | AI-driven threat detection and response for national infrastructure cybersecurity | AI improves speed and accuracy of threat detection and incident response. - Automation reduces dependency on human analysts. AI systems adapt well to evolving and emerging cyber threats across sectors. | - Over-reliance on AI may introduce risks if algorithms fail or misclassify threats. - Requires high-quality, continuously updated datasets. - Integration challenges within legacy infrastructure systems. | - Develop robust fail-safe mechanisms for AI-driven security tools. - Expand datasets for improved model training and threat generalization. - Explore hybrid human-AI operational frameworks for critical infrastructure networks. |
| Mahajan et al. (2023) | Cloud-based security issue detection & response with AI | AI models improve detection accuracy; digital forensics speeds incident triage; proactive threat detection; optimized resource allocation; regulatory compliance. | Adversarial ML risks, compliance complexities, cross-cloud integration issues. | Develop resilient AI-driven strategies, enhance human–machine cooperation, strengthen legal/regulatory frameworks. |
| Sidhu (2023) | Threat Management for Cloud-Based Network Intrusion Detection Systems | NID improves early threat detection and avoids breaches by detecting malicious traffic, illegal access, data leakage, and distributed denial of service. | Handling large-scale cloud traffic; reducing false positives; ensuring continuous monitoring. | Improve scalability of NID, integrate AI for adaptive detection, enhance deployment across diverse cloud architectures. |
| Ademilua & Areghan (2022) | AI-driven cloud security frameworks | Supervised, unsupervised, DL, NLP, RL, federated learning improve threat detection, policy automation, compliance. | Interpretability of AI decisions; evolving complex threats; regulatory alignment. | Build interpretable AI models, expand federated learning, integrate automation for faster response. |
| Oduri (2021) | Real-time monitoring & ML for anomaly detection | AI improves detection rates, reduces false positives, enhances system reliability; validated across cloud platforms. | Data privacy issues; implementation complexity; dependency on training data quality. | Address privacy-preserving ML, optimize AI deployment, explore adaptive models for evolving threats. |
| Patel et al. (2020) | Cloud computing security challenges | Cloud reduces costs, enables distributed work, but introduces security risks from attackers exploiting valuable stored data. | Vulnerabilities in cloud services; unauthorized access; evolving threats. | Strengthen provider–customer trust, develop robust cloud-native security tools. |
| Torkura et al. (2019) | Cloud incident detection & response (SlingShot) | Dynamic snapshotting & automated alert correlation; fast detection (30s) and response (25–90 mins). | Response time disparity (AWS vs GCP); scalability to larger systems. | Optimize real-time response, extend techniques to multi-cloud setups, reduce MTTR further. |
| Al-Mohannadi et al. (2018) | IDS/IPS for cyber-attack handling | IDS/IPS generate large alerts; require human filtering for true positives; need skilled IT staff. | High false positives; reliance on human expertise; alert fatigue. | Develop AI-assisted IDS/IPS to reduce false positives, train workforce, automate incident prioritization. |

## 6. Conclusion and Future Work

AI in enhancing cloud threat identification and response systems, and transitioning to autonomous and adaptive systems instead of reactive and manual systems. Combining supervised, unsupervised, deep learning, reinforcement learning, and natural language processing, AI is aimed at enhancing the accuracy of the detection, reducing the reaction time, and automating remediation. The ability to identify, classify, and neutralize emerging threats in real-time makes AI-driven systems much more resilient to cloud infrastructures. Issues still exist in data privacy, adversarial attacks, model interpretability, and high costs of implementation. Even existing solutions are not free of obstacles to seamless integration into heterogeneous multi-cloud environments, the new reality of enterprise operation. Privacy-preserving AI models, including federated learning and differential privacy should be considered in future research to guarantee high security and compliance with regulations. Creation of interpretable AI will be critical in creating trust and transparency in automated decision-making. Another priority will have been dealing with adversarial machine learning with the use of resilient algorithms and defensive mechanisms. In addition, autonomous response mechanisms can be extended to multi-cloud and hybrid environments to enhance scalability and minimize platform differences in response. Lastly, enhanced human-AI collaboration models will have a role in equalizing automation and human controls alongside ethical, responsible and proactive defence measures against the future cloud ecosystems.